\documentclass[fleqn,usenatbib]{mnras}

\usepackage{mathptmx}

\usepackage[T1]{fontenc}

\DeclareRobustCommand{\VAN}[3]{#2}
\let\VANthebibliography\thebibliography
\def\thebibliography{\DeclareRobustCommand{\VAN}[3]{##3}\VANthebibliography}

\usepackage{graphicx}	
\usepackage{amsmath}	
\usepackage{amssymb}	
\usepackage{braket}
\usepackage{etoc}
\usepackage{mdwlist}
\usepackage{times}
\usepackage{color,soul}
\usepackage{lscape}
\usepackage{inputenc}
\usepackage{wrapfig}
\usepackage{amsmath}
\usepackage{bm}        
\usepackage{mciteplus} 
\usepackage{array,longtable}
\usepackage{adjustbox}
\usepackage{vmargin}   
\usepackage{setspace}
\usepackage{ifthen}
\usepackage{longtable}
\usepackage{rotating}
\usepackage{multirow}
\usepackage[version=3]{mhchem}
\usepackage{lineno}
\usepackage{dblfloatfix}

\title[The methylamine + CN reaction]{A twist on the reaction of the CN radical with methylamine in the interstellar medium: new hints from a state-of-the-art quantum-chemical study}

\author[Puzzarini et al.]{
Cristina Puzzarini,$^{1}$\thanks{E-mail: cristina.puzzarini@unibo.it}
Zoi Salta,$^{2}$
Nicola Tasinato,$^{2}$
Jacopo Lupi,$^{2}$
\newauthor 
Carlo Cavallotti,$^{3}$
and Vincenzo Barone,$^{2}$\thanks{E-mail: vincenzo.barone@sns.it}
\\
$^{1}$Dipartimento di Chimica ``Giacomo Ciamician'', Universit\`a di Bologna, Via F. Selmi 2, 40126 Bologna, Italy\\
$^{2}$Scuola Normale Superiore, Piazza dei Cavalieri 7, I-56126 Pisa, Italy\\
$^{3}$Department of Chemistry, Materials, and Chemical Engineering ``G. Natta'', Politecnico di Milano, I-20131 Milano, Italy
}

\date{Accepted XXX. Received YYY; in original form ZZZ}

\pubyear{2020}

\begin{document}
\label{firstpage}
\pagerange{\pageref{firstpage}--\pageref{lastpage}}
\maketitle


\begin{abstract}
Despite the fact that the majority of current models assume that interstellar complex organic molecules (iCOMs) are formed on dust-grain surfaces, there is some evidence that neutral gas-phase reactions play an important role. In this paper, we investigate the reaction occurring in the gas phase between methylamine (CH$_3$NH$_2$) and the cyano (CN) radical, for which only fragmentary and/or inaccurate results have been reported to date. This case study allows us to point out the pivotal importance of employing quantum-chemical calculations at the state of the art. Since the two major products of the CH$_3$NH$_2$ + CN reaction, namely the CH$_3$NH and CH$_2$NH$_2$ radicals, have not been spectroscopically characterized yet, some effort has been made for filling this gap.
\end{abstract}




\section{Introduction} \label{sec:intro}

Since the early 1960s, the discovery of new molecules in the interstellar medium (ISM) has continued at a nearly steady pace \citep{McGuire_2018}, with the majority of these species being identified thanks to their rotational signatures. Despite the fact that more than 200 molecular species have been detected in the ISM and circumstellar shells, radioastronomical line surveys still present a significant number of unassigned features. Among the molecular species discovered, the so-called interstellar complex organic molecules (iCOMs) have attracted particular attention because most of them can be considered as precursors of biochemical building blocks (see, e.g., \cite{exo,COMs,horst12,C2CS35113G,Saladino-RSC2012_Formamide,SaladinoE2746}). Among iCOMs, the compounds containing the cyano-moiety (CN functional group) play a remarkable role as potential precursors of amino acids, the main constituents of proteins, and nucleobases, the fundamental components of DNA and RNA (see, e.g., \cite{horst12,C2CS35113G,Saladino-RSC2012_Formamide} and references therein). For example, the Strecker synthesis is well-known to lead to the formation of nitrile derivatives that can then evolve to amino acids by hydrolysis of the latter. Among the different variants of this synthetic route, the simplest one involves aminoacetonitrile (NH$_2$CH$_2$CN, AAN) as product, which forms --after its hydrolysis-- glycine. Several theoretical studies have proved that AAN can be indeed obtained by means of the Strecker synthesis (see, e.g., \cite{jp076221,B923439J} and references therein).

While the evidence for molecular complexity in the universe is undisputed, less clear is how chemical evolution takes place and how many molecular species are still hidden from our knowledge. These two challenges are strongly connected. The mechanisms of formation of the detected molecules in the typically cold and (largely) collision free environment of the ISM are often unknown. Their disclosure and understanding would allow for rationalizing the molecular abundances observed in interstellar clouds, but also would help for obtaining a more complete picture of the molecular species possibly existing in the ISM. Indeed, the derivation of feasible reaction pathways might suggest new molecules to be searched for in space, thus requiring their spectroscopic characterization.

In the last decade, grain-surface chemistry has been mostly invoked to explain the formation of molecules in space, the basic idea being that radical species trapped in icy-mantles can react and give rise to a rich chemistry (see, e.g., \cite{Garrod06,Garrod_2008,Oberg_2010,0144235X.2015.1046679,acsearthspacechem.7b00156}). However, the recent observation of complex molecules also in very cold objects (at 10 K only H atoms are able to move on dust particles) has suggested that gas-phase reactions could have been overlooked \citep{Bacmann12,Vasyunin_2013,Vastel_2014,slv009}. Indeed, there are astronomical evidences that gas-phase reactions play a role in the ISM (see, e.g., \cite{slv009,SOLIS-II,0004-637X-854-2-135}). At low temperatures, molecular synthesis through gas-phase chemistry can proceed via either ion-molecule or neutral-neutral reactions, the latter involving at least one radical species. Furthermore, to have a more complete picture, photoionization of gas-phase species and the consequent fragmentation pathways of the resulting cations should also be considered (see, e.g., \cite{BELLILI2015196}).   

Understanding the chemical evolution of an interstellar cloud requires the characterization of thousands of reactions that involve hundreds of species. To develop chemical models able to explain the observed molecular abundances, formation pathways within or upon dust-grain ice mantles as well as in the gas phase should be incorporated in the specific network \citep{Garrod_2011,Garrod_Glycine_HotCores_Alma_ApJ13}. A key point is however to rely on accurate and reliable data. In this scenario, accurate state-of-the-art computational approaches play a fundamental role because they provide a powerful tool for deriving feasible reaction mechanisms as well as accurate predictions of spectroscopic parameters. Concerning reactivity, experimental investigations face difficulties in mimicking the extreme conditions that characterize the ISM (but also planetary atmospheres) in the laboratory, and they often require guidance of theory to be interpreted (see, e.g., \cite{F-H2,jp406842q,jpclett.5b00519,acs.jpca.8b05102,jpclett.7b03439,Yang14471,c9h9}). However, it must be pointed out that accurate determination of reaction mechanisms by theory is at the state of the art in computational chemistry because, at the typical low temperatures of the ISM, rates are extremely sensitive to energetics and barrier heights. In a second step, reaction intermediates and/or products of potential interest to the ISM need to be computationally studied in order to lay the foundation for a subsequent spectroscopic characterization by means of rotational spectroscopy experiments that would enable the knowledge of rotational signatures with the proper accuracy to guide astronomical searches.

The focus of this work is the investigation of the reaction between methylamine (CH$_3$NH$_2$) and the cyano radical (CN) that, despite potentially leading to a wealth of interesting products ({\it vide infra}), has not been yet analysed satisfactorily. On general grounds, the starting point of our approach is the design of a feasible and accessible reactive potential energy surface (PES) leading to the iCOM of interest, with the potential precursors being selected among the molecular species already detected in space. The following step is the investigation of the reactive PES itself with the identification of all stationary points (minima and transition states) along the path using, at this stage, a cost-effective computational model. Usually, different routes toward the sought product or other species can be derived. Among them, only those that can be feasible in the typical conditions of the astronomical environment under consideration will be further investigated. For instance, the ISM is characterized by harsh conditions with extremely cold (down to 10 K) regions where the density is extremely low (of the order of 10$^4$ particles/cm$^3$). In such extreme conditions, accessible chemical routes are those for which all energy barriers lie below the energy of the reactants, that is all transition states should be submerged. Subsequently, for the selected reaction schemes, an effective computational strategy requires the accurate computation of structural, energetic, and vibrational features of all the intermediates and transition states involved. The final steps are: ($i$) the evaluation of the kinetic aspects in order to understand what products can indeed be formed and the corresponding rate; ($ii$) the accurate prediction of the spectroscopic parameters of those products for which such information is still missing, this being the first step toward laboratory measurements.

Coming to the specific subject of our study, the outcomes of new state-of-the-art quantum-chemical computations for the CH$_3$NH$_2$ + CN reactive PES will be presented and compared with the contradictory and incomplete results of previous studies \citep{C7CP05746F,acsearthspacechem.8b00098}, also allowing us to point out the importance of accurate samplings and characterizations. The reaction between the electrophilic CN radical and a molecule with an electron lone pair like methylamine can lead, together with direct H abstraction from the NH$_2$ or CH$_3$ moieties, to a large number of radical species of global formula C$_2$N$_2$H$_5$ that might subsequently evolve in a wealth of products, including  --for example-- AAN and cyanamide (NH$_2$CN):
\begin{eqnarray}
\mathrm{CH_3NH_2 + CN} \;\:\;\; & \rightarrow &  \;\:\;\; \mathrm{CH_2NH_2 + HCN}      \\ \nonumber 
                                           \;\:\;\; & \rightarrow &  \;\:\;\;  \mathrm{CH_3NH + HCN}        \\ \nonumber 
				        	 \;\:\;\; & \rightarrow &  \;\:\;\;  \mathrm{NH_2CH_2CN + H }   \\ \nonumber    
					 \;\:\;\; & \rightarrow &  \;\:\;\;  \mathrm{NH_2CN + CH_3}
\end{eqnarray}
In the present investigation, only the doublet PES has been investigated. In detail, the electronic states of the open-shell reactants and products are: $^2\Sigma^+$ (CN), $^2A'$ (CH$_2$NH$_2$), $^2A''$ (CH$_3$NH), and $^2A''_2$ (CH$_3$). Accordingly, $^2A'$ electronic states have been considered for the open-shell intermediates or transition states leading to CH$_2$NH$_2$ and $^2A''$ electronic states for all the other cases.

Our original interest on this reaction was indeed related to AAN as a possible product, since no gas-phase reactions have been suggested for its interstellar production. As mentioned above, among the potential precursors of amino acids, AAN --which has been detected toward Sagittarius B2(N) \citep{detectionAAN}-- has attracted particular attention due to its involvement in the Strecker synthesis of glycine. Cyanamide is another interesting molecule with a prebiotic potential and, in \cite{C7CP05746F}, it has been claimed to be the product formed in the reaction above at low temperature. However, as will be demonstrated, the reactive PES should be accurately investigated in order to understand which are the potential products in the harsh conditions typical of the ISM. Among them, there are two radical species poorly characterized in the literature \citep{DYKE1989221,Wright19964408,ch2nh2,jpca.6b03516}, namely CH$_2$NH$_2$ and CH$_3$NH, which warrant attention. Interestingly, the CH$_2$NH$_2^+$ cation has been recently investigated by high-resolution rovibrational and pure rotational spectroscopy \citep{C9CP05487A}. Since test computations showed that the attack of the cyanogen radical by the C end is strongly favoured with respect to that by the N atom, only the paths starting from the former type of attack have been further considered. Therefore, the formation of HNC and other isonitrile products has not taken into consideration.
  
The manuscript is organized as follows. First, the essential computational details are provided, with more information given in the Appendix. In the subsequent section, the results are reported and thoroughly discussed: the outcomes of the investigation of the reaction between methylamine and the cyano radical are presented from both a thermochemical and a kinetic point of view. Then, the spectroscopic characterization of the CH$_2$NH$_2$ and CH$_3$NH radicals is deteiled. Finally, concluding remarks are provided.

\section{Computational investigation}

In the following, the computational strategy for accurately investigating the gas-phase methylamine + CN reaction is presented in some details. As mentioned in the Introduction, the spectroscopic technique of choice for the detection of molecular species in space is rotational spectroscopy. For this reason, the details of the corresponding computational spectroscopic characterization are provided.

\subsection{The \ce{CH3NH2 + CN} reaction}

A preliminary scan of the PES of the CH$_3$NH$_2$ + CN reactive system was carried out by using the B3LYP hybrid functional \citep{B-B3LYP,LYP} in conjunction with the double-zeta 6-31+G(d) basis set, which is similar to the level of theory employed for the geometry optimizations performed in the recent investigation of the same system by \cite{C7CP05746F}. Furthermore, such a level is widely used in model chemistry methods (e.g. CBS-QB3 \citep{CBS-QB3,CBS-QB3b}, W1 \citep{ct900260g} or G4 \citep{1.2436888}). Calculations were subsequently refined by means of the double-hybrid B2PLYP functional \citep{Grimme_B2PLYP_JCP06} combined with a modified may-cc-pVTZ basis set \citep{ct800575z,ct9004905,C5CP07386C} ($d$ functions  removed on hydrogens), in the following referred to as may$^{\prime}$-cc-pVTZ. This level of theory has been demonstrated to perform well at a reduced computational cost \citep{CHEM201606014,Melli-AstrophysJ-2018,NE-TMA,anie.201906977,C9CP06768J,BOUSSESSI2020127886}. Since semi-local density functional approximations fail to correctly describe the long-range London dispersion interactions \citep{grimme_wiley2011}, these effects were taken into account by the Grimme's D3 scheme \citep{D3} employing the Becke-Johnson (BJ) damping function \citep{jcc.21759}. To check the nature of all stationary points, the corresponding Hessian matrices were evaluated. Saddle points were assigned to reaction paths by using intrinsic reaction coordinate (IRC) calculations \citep{ar00072a001} for the identification of reactants and products.

To check possible structural effects on the energetics, in addition to the B2PLYP-D3(BJ)/may$^{\prime}$-cc-pVTZ level of theory, the so-called ``cheap'' geometry scheme \citep{ura,Thiouracil_B0_PCCP13,Puzzarini_JPCL2014_GLYdipept,thioura-h2o} has been considered for optimizing the equilibrium geometries of the reactants, some intermediates and products. This approach, which is described in the details in Appendix \ref{appendix:cheap}, starts from the coupled-cluster (CC) method including full account of single and double excitations and a perturbative estimate of triple excitations, CCSD(T) \citep{Raghavachari-CPL1989_CCSD_T}, in conjunction with a triple-zeta quality basis set and incorporates the extrapolation to the complete basis set (CBS) limit and the core-correlation contribution by making use of M\o ller-Plesset theory to second order, MP2 \citep{MP34}. According to the literature on this topic (see, e.g., \cite{Barone-PCCP2013_Glycine_conformers2,QUA25202,account-structure}), its accuracy is expected to be of about 0.001-0.002 \AA\ for bond distances and around 0.1-0.2 deg. for angles.   

Subsequently, the energetics of all stationary points was accurately determined by applying different composite schemes:
\begin{enumerate}
	\item The CBS-QB3 model chemistry. This scheme employs a CC ansatz in conjunction with complete basis set extrapolation and uses B3LYP/6-31G(d) geometry optimizations and zero-point energies. Empirical corrections are also introduced in this model. For details, the reader is referred to \cite{CBS-QB3,CBS-QB3b}.
	\item CCSD(T)/VTZ. Since the CCSD(T) method is referred to as the ``gold standard'' for accurate quantum-chemical calculations, it is often used, in conjunction with the cc-pVTZ basis set \citep{Dunning-JCP1989_cc-pVxZ} and within the frozen-core (fc) approximation, in the investigation of reactive PESs.
	\item The ``CCSD(T)/CBS+CV'' composite scheme. This is entirely based on CCSD(T) calculations and accounts for the extrapolation to the CBS limit and for core-correlation effects (see, e.g., \cite{Miriam2,hosop,Barone-PCCP2013_Glycine_conformers2,Puzzarini_Acrolein_Re_IR_JPCA14}). It is described in detail in Appendix \ref{appendix:cbscv}.
	\item The approach denoted as ``HEAT-like''. Starting from the CCSD(T)/CBS+CV approach, this composite scheme improves it by incorporating the contributions due to the full treatment of triple excitations and a perturbative treatment of quadruples as well as diagonal Born-Oppenheimer and relativistic corrections (see, e.g., \cite{heat,heat2,heat3,hosop}). The methodology is described in Appendix \ref{appendix:heat}.
\end{enumerate}

Except for a few, selected stationary points, all single-point energy calculations at the fc-CCSD(T)/cc-pVTZ and CCSD(T)/CBS+CV levels as well as using the ``HEAT-like'' model were performed on top of B2PLYP-D3(BJ)/may$^{\prime}$-cc-pVTZ optimized geometries.   
For open-shell species, correlated calculations have been carried out by using restricted open-shell Hartree-Fock (ROHF) reference wavefunctions. 
In the last step of the PES characterization, electronic energies need to be augmented by zero-point vibrational energy (ZPE) corrections. The latter have been obtained using vibrational perturbation theory to second order (VPT2; \cite{Bloino-JCTC2012_HDCPT2}) applied to B2PLYP-D3(BJ)/may$^{\prime}$-cc-pVTZ anharmonic force fields.

CBS-QB3 calculations as well as DFT geometry optimizations and force field computations were performed with the Gaussian 16 quantum-chemical software \citep{G16C01}. Calculations for the ``cheap'', CCSD(T)/CBS+CV, and ``HEAT-like'' schemes were carried out using the quantum-chemical CFOUR program package \citep{CFour}, except those including quadruple excitations which have been performed with the MRCC code \citep{mrcc} interfaced to CFOUR.  

\subsection{Kinetic models}

Global rate constants, for both addition-abstraction on the NH$_2$ group and abstraction of H from the methyl group by CN, were calculated using a master equation (ME) approach based on ab initio transition state theory (AITSTME). For this purpose, the MESS software was used (available at https://github.com/PACChem/MESS), which features the strategies detailed in \cite{georgievskii2013reformulation}. Rate constants for the elementary reactions passing through a saddle point were computed using conventional transition state theory (TST), also accounting for tunneling effects by means of the Eckart model \citep{eckart1930penetration}. Rate constants for barrierless elementary reactions were evaluated using a two-transition-states model as implemented in MESS, with the two transition states describing the long range and short range dynamic bottlenecks usually found in barrierless reactions. Microcanonical rate constants for each transition state were determined using variable reaction coordinate transition state theory (VRC-TST). Long range VRC-TST calculations were performed on spherical dividing surfaces sampling the PES as a function of distances comprised between 20.0 and 9.0 $a_0$, measured with respect to two pivot points positioned in the centers of mass of the reactive fragments. Short range VRC-TST calculations were performed using multifaceted dividing surfaces placing two pivot points on the CH$_3$NH$_2$ reacting atoms, symmetrically displaced along the direction of the breaking/forming bond (0.01-0.3 $a_0$), and a single pivot point centered on the CN carbon atom. The sampled short range distances were comprised between 8.5 and 3.5 $a_0$. The interaction potential was computed at the CASPT2 level \citep{caspt2_1,caspt2_2,caspt_3}, as described in Appendix \ref{appendix:vrc}. In the case of H abstraction from methyl, restrained geometry optimizations showed that abstraction of the H atom positioned on the CH$_3$NH$_2$ symmetry plane is largely favored over that of out-of-plane hydrogens. VRC-TST calculations were thus performed only for this pathway. For this reaction, it is difficult to differentiate the reacting flux from that leading to addition to the NH$_2$ group or to H abstraction from the other methyl hydrogens. For this reason, a fictitious repulsive potential was added to the CASPT2 potential when the distance between the CN carbon atom and the nitrogen atom or the two out-of-plane H atoms of methyl is smaller than that between CN and the abstracted H atom. This effectively allows to raise the energy of the states leading to the competitive reaction pathways, thus decreasing their contribution to the density of states of the transition state for the investigated reaction pathway. The calculated microcanonical reactive fluxes were multiplied by a flat 0.9 factor to correct for recrossing of the dividing surface. VRC-TST calculations were performed using the VaReCoF software \citep{varecof}, generating the necessary input files through EStokTP \citep{cavallotti2018estoktp}. CASPT2 calculations were carried out using the MOLPRO program \cite{MOLPRO_brief}.

\begin{figure*}
\centering
\includegraphics[width=2.0\columnwidth]{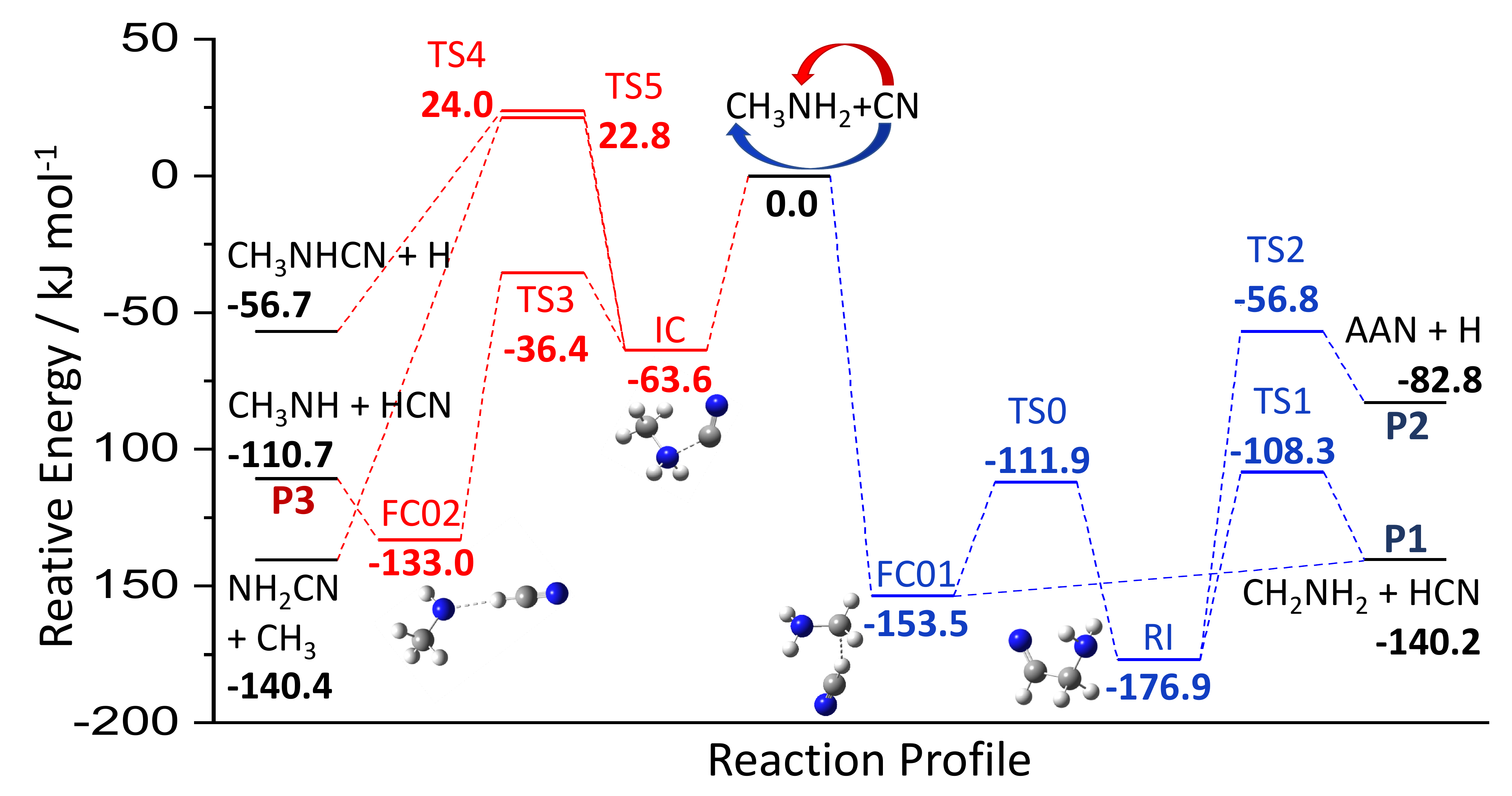}
\caption{Reaction mechanism for the attack of CN to the N moiety of methylamine in red and for the abstraction of H from the methyl group by CN in blue. ``HEAT-like'' energies augmented by anharmonic ZPE corrections. \label{tot-path}}
\end{figure*}

\subsection{Spectroscopic characterization}

The rationalization of rotational spectra is made in terms of an effective rotational Hamiltonian, whose leading terms are the rotational constants. For the vibrational ground state, according to VPT2 \citep{Mills-chap1972_VibRot_Struct}, they can be written as:

\begin{equation}
B_0^i = B_e^i ({\mathrm{best}}) + \Delta B_{vib}^i ({\mathrm{B2}}) \; ,
\end{equation}
where $B_e^i$ denotes the equilibrium rotational constant with respect to the $i$-th inertial axis ($i$ = $a,b,c$, so that $B_e^a$ = $A_e$), and $\Delta B_{vib}^i$ the corresponding vibrational correction. Since the $B_e$'s only depend on the equilibrium structure, the latter was obtained using the CCSD(T)/CBS+CV composite scheme further improved by accounting for the full treatment of triple and quadruple excitations \citep{Miriam1,Miriam2}, as explained in Appendix \ref{appendix:spectro}. Indeed, although B2PLYP-D3 geometry optimizations meet well the accuracy requirements for energy evaluations, predictions of equilibrium rotational constants need a much higher precision in equilibrium structure determinations. For this reason, we resorted to the composite scheme mentioned above.
$\Delta B_{vib}$ corrections were computed from B2PLYP-D3(BJ)/may$^{\prime}$-cc-pVTZ anharmonic force fields by applying the VPT2 implementation available in Gaussian \citep{Barone-JCP2005_VPT2}. A reduced dimensionality approach \citep{qua.23224} has been employed for both radicals in order to properly treat the internal methyl rotation in CH$_3$NH and the NH$_2$ inversion in CH$_2$NH$_2$. Anharmonic force-field calculations also provided, as a byproduct, the quartic and sextic centrifugal-distortion constants.

To complete the rotational spectroscopy characterization, the electron spin-rotation tensor together with the hyperfine coupling and nitrogen quadrupole coupling constants need to be computed, with all computational details provided in Appendix \ref{appendix:spectro}. The electron spin-rotation interaction originates from the coupling between the rotational angular momentum and the electron spin (thus being a second-order property) and the corresponding tensor has been evaluated at the CCSD(T)/cc-pCVQZ level of theory, with all electrons correlated. Hyperfine coupling and nitrogen quadrupole coupling constants are instead first-order properties and have been calculated at the CCSD(T)/aug-cc-pCVQZ level (all electrons correlated), with the basis sets for the hydrogen atoms being modified as explained in the Appendix \ref{appendix:spectro}. Equilibrium parameters were finally corrected for vibrational effects within the VPT2 approach \citep{Barone-JCP2005_VPT2,chemrev.9b00007} at the B2PLYP-D3(BJ)/may$^{\prime}$-cc-pVTZ level of theory. The only exception is the electron spin-rotation tensor, for which vibrational corrections have been computed at the B3LYP/6-31+G(d) level due to the lack of the required B2PLYP implementation.

\section{Results and discussion}

First of all, a detailed analysis of the gas-phase reaction between methylamine and the cyano radical is reported from thermochemical and kinetic points of view. From this, the spectroscopic interest on the CH$_2$NH$_2$ and CH$_3$NH radicals will become clear and detailed afterward.

\subsection{The \ce{CH3NH2 + CN} reaction}

\begin{figure}
\centering
\includegraphics[width=1\columnwidth]{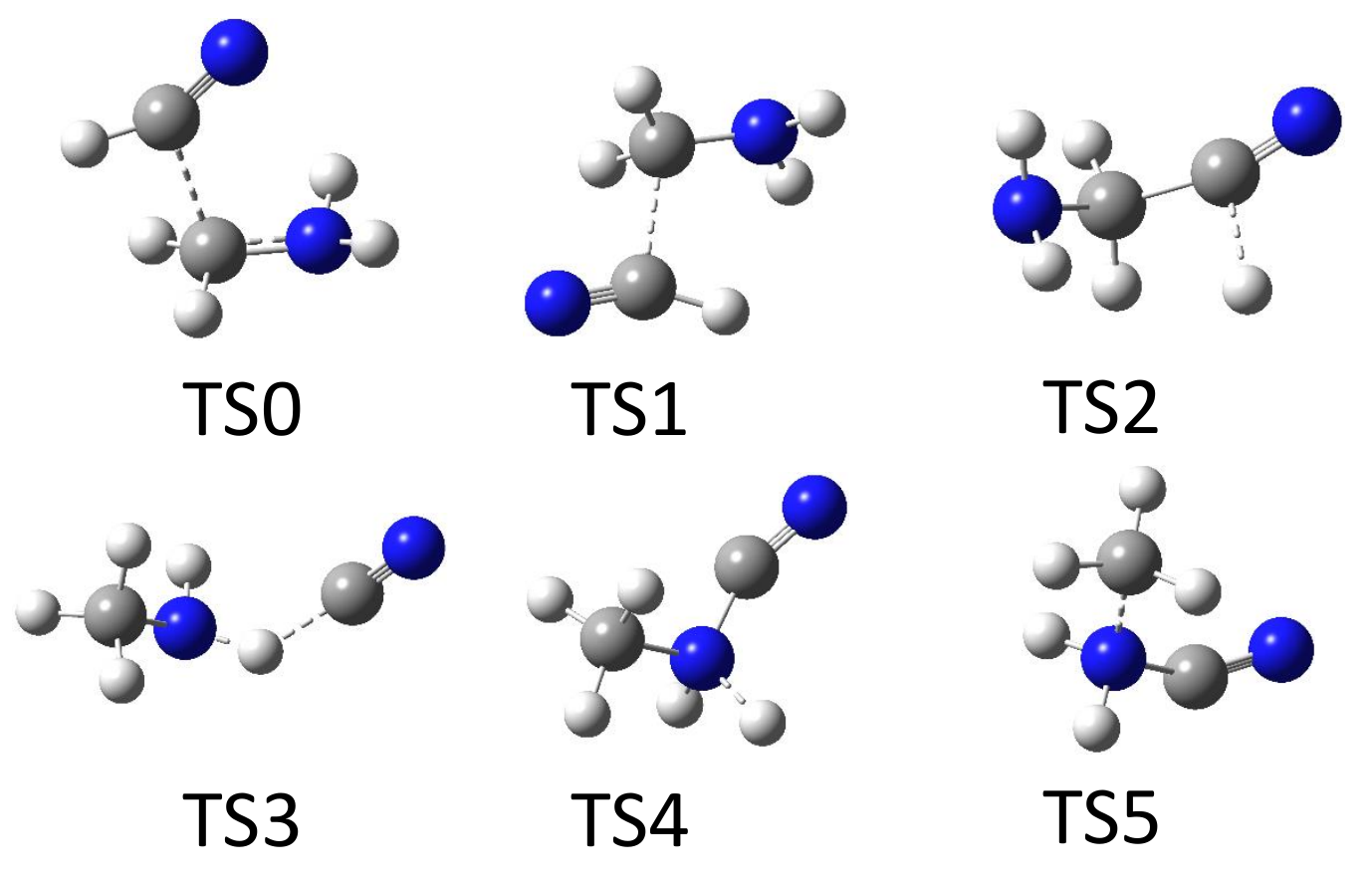}
\caption{Transition states of the methylamine + CN reaction. \label{figTS}}
\end{figure}

Focusing on the CH$_3$NH$_2$ + CN reaction, it has to be noted that none of the previous works \citep{C7CP05746F,acsearthspacechem.8b00098} provided a complete picture of the general mechanism, which --instead-- has been thoroughly investigated in the present study. This is summarized in Figure~\ref{tot-path}: paths in red are the results of the attack of CN to the N moiety of methylamine, those in blue of the abstraction of H from the methyl group by CN (the relative electronic energies, obtained at different computational levels, and the corresponding ZPEs are detailed in Table \ref{tab}). For convenience of the reader, the structures of the transition states involved in the reaction mechanisms displayed in Figure~\ref{tot-path} are shown in Figure~\ref{figTS}. 

To discuss the details of these two pathways, we consider the prototypical additions of the CN radical to ammonia or methane, which have been thoroughly investigated in \cite{B908416A} and \cite{C7CP03499G}, respectively. As a matter of fact, the ``FC01 route'' resembles the addition of CN to ammonia and the ``FC02 route'' that to methane. However, methylamine shows a cooperative effect of the amine and methyl group leading to some differences.

\begin{table*}
\caption{Relative energies, at different levels of theory, and ZPE corrections for the CH$_3$NH$_2$ + CN reaction.\textsuperscript{\emph{a}} Values in kJ mol$^{-1}$.}\label{tab}
\begin{tabular}{llcccc|cc}
\hline
\hline
Label     & Chemical Formula         & HEAT-like\textsuperscript{\emph{b}} & CBS+CV\textsuperscript{\emph{c}} & CCSD(T)/VTZ\textsuperscript{\emph{d}} & CBS-QB3 & anharm-ZPE\textsuperscript{\emph{e}} & harm-ZPE\textsuperscript{\emph{f}}\\
\hline
Reactants & CH$_3$NH$_2$+CN          &    0.0       &    0.0     &    0.0          &    0.0  &   0.0     &   0.0  \\ 
FC01      & H$_2$N-H$_2$C$\cdots$HCN & -150.0       & -153.0     & -147.0          & -151.6  &  -3.5    &  -2.9   \\       
          &                          & (-150.1)     & (-153.0)   & (-147.2)       & & &  \\                                      
RI        & H$_2$NH$_2$CCNH          & -187.4       & -189.7     & -183.2          & -191.0  &  10.5    &  10.3   \\       
IC      & H$_3$CH$_2$N$\cdots$CN   &  -71.5       &  -74.3     &  -70.0          &  -76.4  &   7.9    &   8.0   \\       
FC02      & H$_3$CHN$\cdots$CHN      & -127.6       & -130.3     & -132.8          & -131.2  &  -5.4    &  -4.7   \\ 
P1        & CH$_2$NH$_2$ + HCN       & -133.4       & -136.5     & -129.3          & -135.1  &  -6.8    &  -7.0   \\       
          &                          & (-133.4)     & (-136.4)   & (-129.5)     & & &    \\                                     
P2        & NH$_2$CH$_2$CN + H       &  -68.3       &  -71.4     &  -67.4          &  -71.3  &  -14.5    & -14.3   \\       
P3        & CH$_3$NH + HCN           & -100.9       & -103.6     & -103.8          & -103.5  &  -9.8    & -10.1   \\       
P4        & CH$_3$NHCN + H           &  -41.0       &  -44.4     &  -40.4          &  -47.5  &  -15.7    & -16.0   \\       
P5        & NH$_2$CN + CH$_3$        & -125.1       & -128.5     & -124.8          & -125.2  &  -15.3    & -16.1   \\ 
TS0       & FC01 $\rightarrow$ RI    & -112.8       & -113.5     & -105.2          & -117.5  &   0.9    &   0.9   \\       
TS1       & RI   $\rightarrow$ P1    & -108.0       & -108.7     & -100.0          & -112.9  &  -0.3    &  -0.1   \\       
TS2       & RI   $\rightarrow$ P2    &  -47.3       &  -49.6     &  -42.3          &  -45.0  &  -9.5    &  -9.2   \\       
TS3       & IC $\rightarrow$ FC02  &  -31.3       &  -30.5     &  -29.7          &  -26.8  &  -5.1    &  -4.6   \\       
TS4       & IC $\rightarrow$ P4    &   31.9       &   28.4     &   39.9          &   31.0  &  -7.9    &  -6.7   \\       
TS5       & IC $\rightarrow$ P5    &   22.0      &   18.8     &   27.6          &   13.8  &   0.9    &   1.3   \\ 
MAX\textsuperscript{\emph{g}}   &                          &              &    3.5     &    8.0          &   8.1   &  & 1.2 \\          
MAE\textsuperscript{\emph{h}}   &                          &              &    2.5     &    3.9          &   3.5   &  & 0.4 \\
\hline
\end{tabular}
\scriptsize{
$^{a}$ Equilibrium structures at the B2PLYP-D3(BJ)/may$^{\prime}$-cc-pVTZ level. Values within parentheses have been obtained using ``cheap'' geometries as reference. \\
$^{b}$ CCSD(T)/CBS+CV+fT+pQ+DBOC+rel level of theory, as explained in the Appendix. 
$^{c}$ CCSD(T)/CBS+CV level of theory, as explained in the Appendix. \\
$^{d}$ fc-CCSD(T)/cc-pVTZ level of theory. 
$^{e}$ Anharmonic ZPEs from VPT2 calculations based on the B2PLYP-D3(BJ)/may$^{\prime}$-cc-pVTZ anharmonic force field. \\
$^{f}$ Harmonic ZPEs at the B2PLYP-D3(BJ)/may$^{\prime}$-cc-pVTZ level. 
$^{g}$ Maximum unsigned deviation with respect to the HEAT-like results. For ZPE, harmonic with respect to anharmonic corrections. 
$^{h}$ Mean absolute error deviation with respect to the HEAT-like results. For ZPE, harmonic with respect to anharmonic corrections.
}
\end{table*}

According to a recent quantum-chemical study \citep{B908416A}, the reaction between the CN radical and NH$_3$ does not proceed significantly toward the H$_2$NCN + H products, at least at low temperatures. The reaction path leading to the formation of the HCN + NH$_2$ products proceeds via a potential well associated with a pre-reactive complex, NC$\cdots$NH$_3$, which evolves in an inner transition state (with the energy barrier being submerged) that, passing through a NCH$\cdots$NH$_2$ intermediate, forms the HCN + NH$_2$  products. The corresponding path for the reaction between methylamine and CN is analogous: it goes through the pre-reactive complex IC, the submerged transition state TS3, and then the FC02 complex, to lead to the HCN + CH$_3$NH products (P3). The major difference when moving from NH$_3$ to CH$_3$NH$_2$ is the stabilization of the two intermediates by about 30 kJ mol$^{-1}$.

The formation of NH$_2$CN and CH$_3$ (P5) as products from IC was considered in \cite{C7CP05746F} as the most probable route based on the hypothesis that the energy barrier due to TS5 was strongly overestimated by their CCSD(T) computations. However, our state-of-the-art computations show that this reaction channel is closed at low temperatures, since TS5 lies about 20 kJ mol$^{-1}$ above the reactants. The same applies for the production of CH$_3$NHCN through elimination of H (P4), which involves a transition state (TS4) about 30 kJ mol$^{-1}$ above the reactants. In summary, upon addition of CN to the nitrogen atom of methylamine, the only open channel is the formation of HCN + CH$_3$NH, with the transition state (TS3) being about 30 kJ mol$^{-1}$ below the reactants.  

A second possible reaction channel corresponds to the attack to the methyl end of methylamine, which resembles the attack of CN to methane. In the case of CH$_4$, several studies agree in suggesting the following mechanism (see, e.g., \cite{C7CP03499G}):
\begin{eqnarray*}
\mathrm{CN + CH_4} \rightleftharpoons \mathrm{RC} \rightarrow \mathrm{TS} \rightarrow \mathrm{PC} \rightleftharpoons \mathrm{HCN + CH_3}  \; ,
\end{eqnarray*}      
where RC and PC are, respectively, the reactant and product complexes, and TS is the transition state connecting them. In the case of methylamine, the assistance by the nitrogen atom makes the RC-TS-PC part collapse into the FC01 complex, which leads to HCN + NH$_2$CH$_2$ (P1) without any potential energy barrier. Although it could seem surprising that the TS barrier of about 10 kJ mol$^{-1}$ reported in \cite{C7CP03499G} for addition to methane completely disappears for methylamine, both DFT and CASPT2 computations (vide infra) agree on the barrierless nature of the latter reaction step and preliminary accurate computations on the methane reaction suggest that the previously reported barrier could be strongly overestimated. In any case, HCN should be formed together with NH$_2$CH$_2$ and its CH$_3$NH isomer. However, another path is possible, which has never been investigated before. In fact, FC01 can rearrange to the more stable RI species through the submerged transition state TS0, which lies about 110 kJ$^{-1}$ mol below the reactants. RI can, in turn, lead either to CH$_2$NH$_2$ + HCN through the submerged transition state TS1 or to AAN + H (P2) through the submerged transition state TS2. Although the formation of aminoacetonitrile appears quite disfavored, at least at low temperatures, the process is a quite simple mechanism and the reaction channel is also open under the conditions of the ISM.

\begin{table*}
 \centering
\caption{Intermolecular distance for the FC01, IC and FC02 complexes computed at different levels of theory.}\label{tab2} 
\begin{tabular}{l|ccc} \hline  \hline      
Level of theory & FC01           & IC                 & FC02 \\
                         & H$\cdots$C & N$\cdots$C & N$\cdots$H \\ 
                         &\multicolumn{1}{c}{\includegraphics[width=1.7cm]{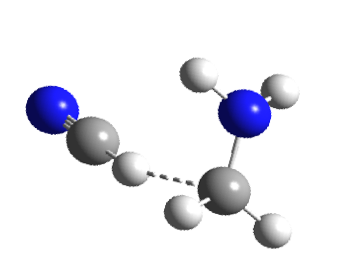}} &\multicolumn{1}{c}{\includegraphics[width=1.6cm]{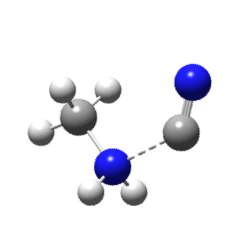}} &\multicolumn{1}{c}{\includegraphics[width=2.3cm]{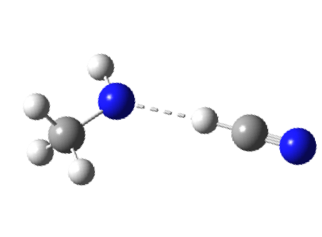}} \\  \hline  
``cheap''                     & 2.329   & 1.992 & 2.093 \\
B2PLYP-D3(BJ)\textsuperscript{\emph{a}} & 2.291   & 2.043 & 2.076 \\
B3LYP\textsuperscript{\emph{b}}                & 2.300   & 2.069 & 2.078 \\
B3LYP-D3(BJ)\textsuperscript{\emph{b}}    & 2.232   & 2.054 & 2.040 \\  \hline  
 \end{tabular}
 
\textsuperscript{\emph{a}}In conjunction with the may$^{\prime}$-cc-pVTZ basis set. \\
\textsuperscript{\emph{b}}In conjunction with the 6-31+G(d) basis set. 
\end{table*}

\subsubsection{Notes on the accuracy of results}

After the discussion of the reaction mechanism, some remarks about the accuracy of structural and energetic determinations are deserved. First of all, we note that, according to the literature at our disposal (see, e.g., \cite{Penocchio_rSE_B2PLYP_JCTC15,Biczysko-WIRES2018_Astrochem,BOUSSESSI2020127886}), for standard closed-shell molecules, B2PLYP-D3(BJ)/may$^{\prime}$-cc-pVTZ structures are predicted to have an accuracy of about 0.002-0.003 \AA\ for bond lengths and about 0.2-0.5 degrees for angles. Moving to open-shell systems, as is the case for transition states, intermediates and some products, a slight worsening of such an accuracy might occur. Nonetheless, uncertainties of this order of magnitude on geometries lead to negligible errors in computed relative stabilities and activation barriers. Even the B3LYP/6-31+G(d) computational level, which is widely employed, e.g., in the CBS-QB3 scheme or in combination with CCSD(T)/cc-pVTZ energies, can be usually considered sufficiently reliable. The situation is different when weakly bonded systems are involved and, unfortunately, systematic studies are not yet available for open-shell systems. For this reason, for the FC01, IC and FC02 complexes, we have checked the accuracy of B3LYP and B2PLYP-D3(BJ) structures by resorting to the so-called ``cheap'' geometry approach as reference. The intermolecular distances of the adducts mentioned above are collected in Table~\ref{tab2}, with their structures also displayed in Figure~\ref{tot-path}.

Table~\ref{tab2} shows that none of the DFT approaches can be considered fully reliable for intermolecular distances. In particular, for B3LYP structures, a clear conclusion cannot be drawn because, for two cases out of three, the agreement is rather good, but for IC the disagreement is relevant. Inclusion of dispersion corrections (D3) always decreases the distances, thus leading to either improvement or worsening. Even B2PLYP-D3(BJ) results, which are our customary standard, show discrepancies of up to 0.05 \AA\ from the reference values. Therefore, the investigation of the effect of such a disagreement on the energetics was deserved. We have computed the relative energy of FC01 and P1 with respect to reactants employing the ``cheap'' structures. The results, provided within parentheses in Table~\ref{tab}, show that --for all levels of theory considered-- only negligible differences (0.1 kJ mol$^{-1}$)  are obtained when using ``cheap'' or B2PLYP-D3(BJ) geometries, with the error essentially vanishing for covalently bonded systems, i.e. P1. As a consequence, B2PLYP-D3(BJ)/may$^{\prime}$-cc-pVTZ structures have been confidently employed in our study.

Moving to the accuracy of energetics, according to the results of Table~\ref{tab}, only CBS+CV values show a maximum error within the so-called chemical accuracy (i.e. 1 kcal mol$^{-1}$, $\sim$4 kJ mol$^{-1}$) with respect to the ``HEAT-like'' reference numbers. 
The maximum error more than doubles moving to the CBS-QB3 and CCSD(T)/cc-pVTZ models, although their MAE remains within the chemical accuracy, provided that restricted open-shell (and not unrestricted) reference wave functions are used.
In this respect, the difference between anharmonic and harmonic ZPEs (MAX = 1.2 kJ mol$^{-1}$, MAE = 0.4 kJ mol$^{-1}$) suggests that the more costly VPT2 computations are warranted only in connection with ``HEAT-like'' or similar composite models.

\begin{figure}
    \centering
    \includegraphics[width=1.1\columnwidth]{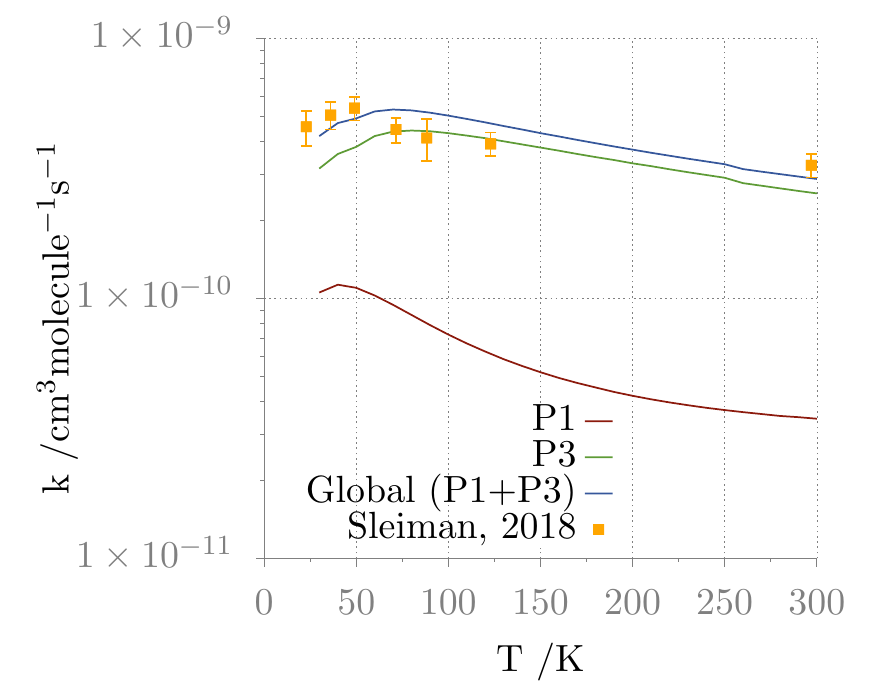}
    \caption{Global rate constants (blue line) and channel specific rate constants leading to the formation of P1 and P3 compared with literature experimental data \citep{C7CP05746F}.}
    \label{fig:rates}
\end{figure}

\begin{figure}
\centering
   \includegraphics[width=1\columnwidth]{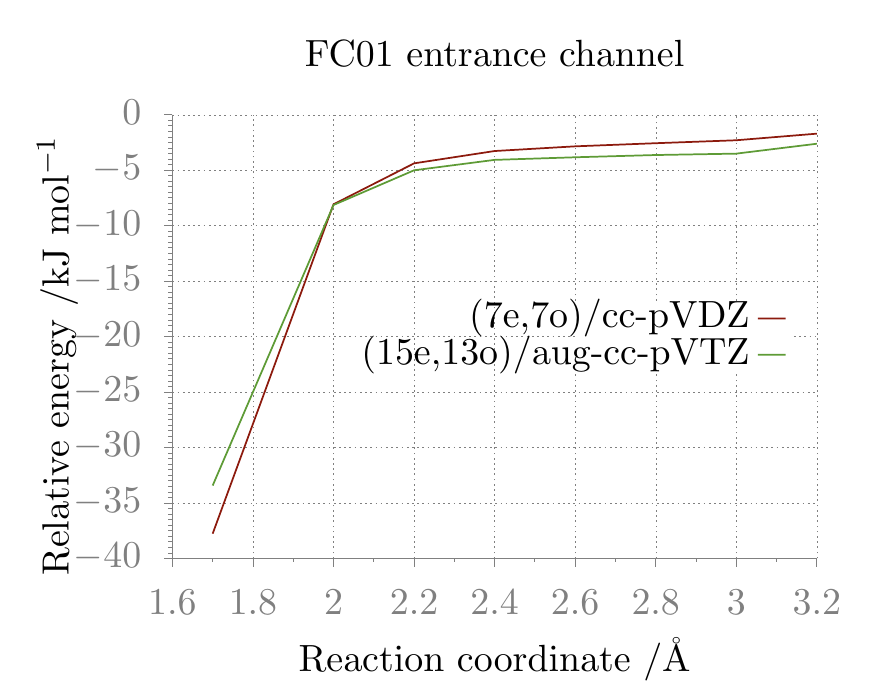}
%
   \includegraphics[width=1\columnwidth]{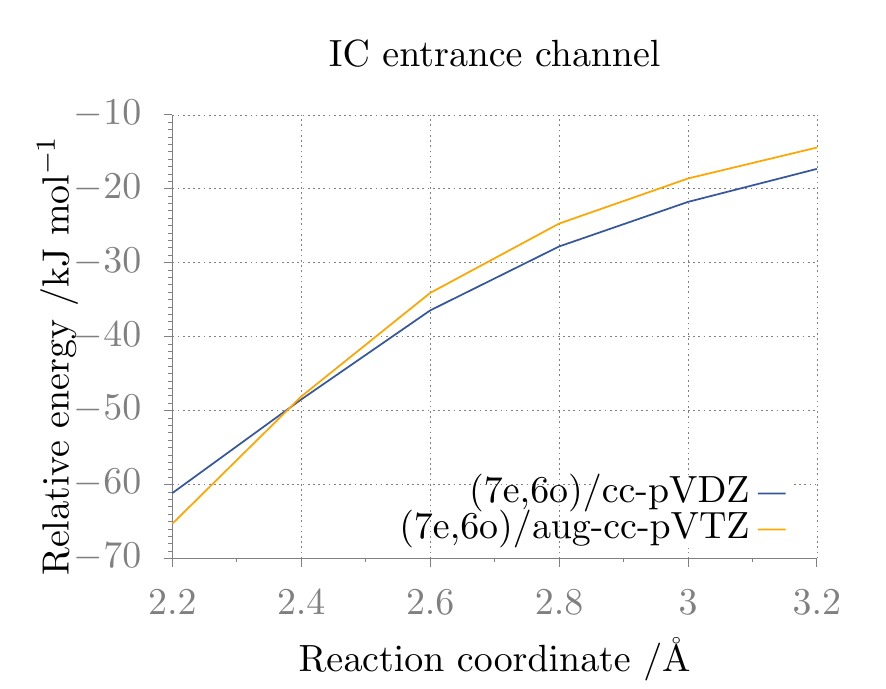}
%
\caption{CASPT2 interaction potentials between CN and CH$_3$NH$_2$ calculated using: (a) the (7e,7o) active space with the cc-pVDZ basis set for constrained optimizations as a function of the \ce{N\bf{C}\bond{...}\bf{H}\bond{-}CH2NH2} distance and the (15e,13o) active space with the aug-cc-pVTZ basis set on (7e,7o) geometries; (b) the (7e,6o) active space with the cc-pVDZ basis set for constrained optimizations as a function of the \ce{N\bf{C}\bond{...}\bf{N}H2CH3} distance and with the aug-cc-pVTZ basis set on cc-pVDZ geometries.}
\end{figure}

\subsection{Rate constants}

\begin{table}
 \centering
\caption{Product-formation rate constants (in cm$^3$ molecule$^{-1}$ s$^{-1}$) at 1 bar as a function of the temperature.}\label{tabella:rates}
\begin{tabular}{cccc}
\hline \hline
 &\multicolumn{2}{c}{Abstraction from CH$_3$} & \multicolumn{1}{c}{Addition to NH$_2$} \\
\cline{2-4}
$T$ /K & P1               & P2               & P3               \\
\hline
30&  1.05$\times$10$^{-10}$ & 1.89$\times$10$^{-17}$   & 3.16$\times$10$^{-10}$  \\
40&  1.13$\times$10$^{-10}$ & 2.05$\times$10$^{-17}$   & 3.83$\times$10$^{-10}$  \\
50&  1.10$\times$10$^{-10}$ & 2.03$\times$10$^{-17}$   & 4.21$\times$10$^{-10}$  \\
60&  1.03$\times$10$^{-10}$ & 1.93$\times$10$^{-17}$   & 4.38$\times$10$^{-10}$  \\
70&  9.44$\times$10$^{-11}$ & 1.80$\times$10$^{-17}$   & 4.42$\times$10$^{-10}$  \\
80&  8.64$\times$10$^{-11}$ & 1.67$\times$10$^{-17}$   & 4.39$\times$10$^{-10}$  \\
90&  7.90$\times$10$^{-11}$ & 1.56$\times$10$^{-17}$   & 4.32$\times$10$^{-10}$  \\
100& 7.27$\times$10$^{-11}$ & 1.46$\times$10$^{-17}$   & 4.23$\times$10$^{-10}$  \\
300& 3.45$\times$10$^{-11}$ & 1.22$\times$10$^{-17}$   & 2.54$\times$10$^{-10}$  \\ \hline
\end{tabular}
\end{table}

Global and channel specific rate constants were computed over the PES shown in Figure~\ref{tot-path} solving the multi-well one-dimensional master equation using the chemically significant eigenvalues (CSEs) method within the Rice-Ramsperger-Kassel-Marcus (RRKM) approximation, as detailed by \cite{miller2006master}. The collisional energy transfer probability is described using the exponential down model \citep{tardy1966collisional} with a temperature dependent $\braket{\Delta E}_{\mathrm{down}}$ of  $260\times(T/298)^{0.875}$ cm$^{-1}$ in an argon bath gas. The rate coefficients for the formation of the P1, P2, and P3 products were computed in the 30-300 K temperature range and at pressure of 0.001-1 bar, the results being collected in Table \ref{tabella:rates}. The corresponding temperature dependence plots are shown in Figure \ref{fig:rates} for the main reaction channels, where they are also compared with experimental data. For these calculations, the “HEAT-like" energies were employed for reactions involving a non negligible transition state, while rate constants of the barrierless channels were computed using VRC-TST. Within the temperature interval considered, the fastest reaction channel is always addition-abstraction to the NH$_2$ group, though the relevance of H abstraction from methyl increases as the temperature decreases. The agreement with experimental data is quite good in the considered temperature range. In order to understand the reason behind the different reactivity of the methyl and amino groups with CN, both characterized by barrierless reaction pathways, it is useful to compare the calculated CASPT2 interaction potentials, reported in Figures \ref{fig:fc01_en} and \ref{fig:ic_en}. It can thus be noted that, for equal \ce{NC\bond{...}H} distances, the \ce{N\bf{C}\bond{...}\bf{C}H3NH2} interaction is significantly more attractive than the \ce{N\bf{C}\bond{...}\bf{H}\bond{-}CH2NH2} interaction, thus leading to the predominance of the P3 reaction channel. As the temperature decreases below 50 K, long range interactions, of similar entity for both attacks, become dominant and the channel P1 branching fraction increases up to 25\%, thus giving an important contribution to the overall reactivity. Pressure does not influence the reaction rate, as the reactants always proceed to form the products without experiencing significant collisional stabilization in the investigated pressure range. 

Finally, a comment is deserved on the much simpler phase space theory (PST), which is usually employed in kinetic studies related to astrochemical processes (see e.g., \citet{0004-637X-810-2-111,BALUCANI201830,10.1093/mnras/sty2903}). Indeed, PST provides a useful, and easy to be implemented, reference theory for barrierless reactions. The basic assumption is that the interaction between two reacting fragments is isotropic and does not affect the internal fragment motions \citep{doi:10.1021/cr050205w}, such an approximation being often valid for low-temperature phenomena, as those occurring in the ISM. In the present case, PST results obtained fitting B2PLYP-D3(BJ) energies as an inverse function of the distance (R) between the fragment centers of mass are in fair agreement with the VRC-TST results for the path leading to IC, but off by about one order of magnitude for the path leading to FC01. This trend is explained by the curves shown in Figures \ref{fig:ic_en} and \ref{fig:fc01_en}: a smooth R$^{-6}$ function well describes the former path, whereas this is not the case for the latter.

\subsection{Spectroscopic characterization of the \ce{CH2NH2} and \ce{CH3NH} radicals}

\begin{figure}[b]
\centering
\includegraphics[width=1.1\columnwidth]{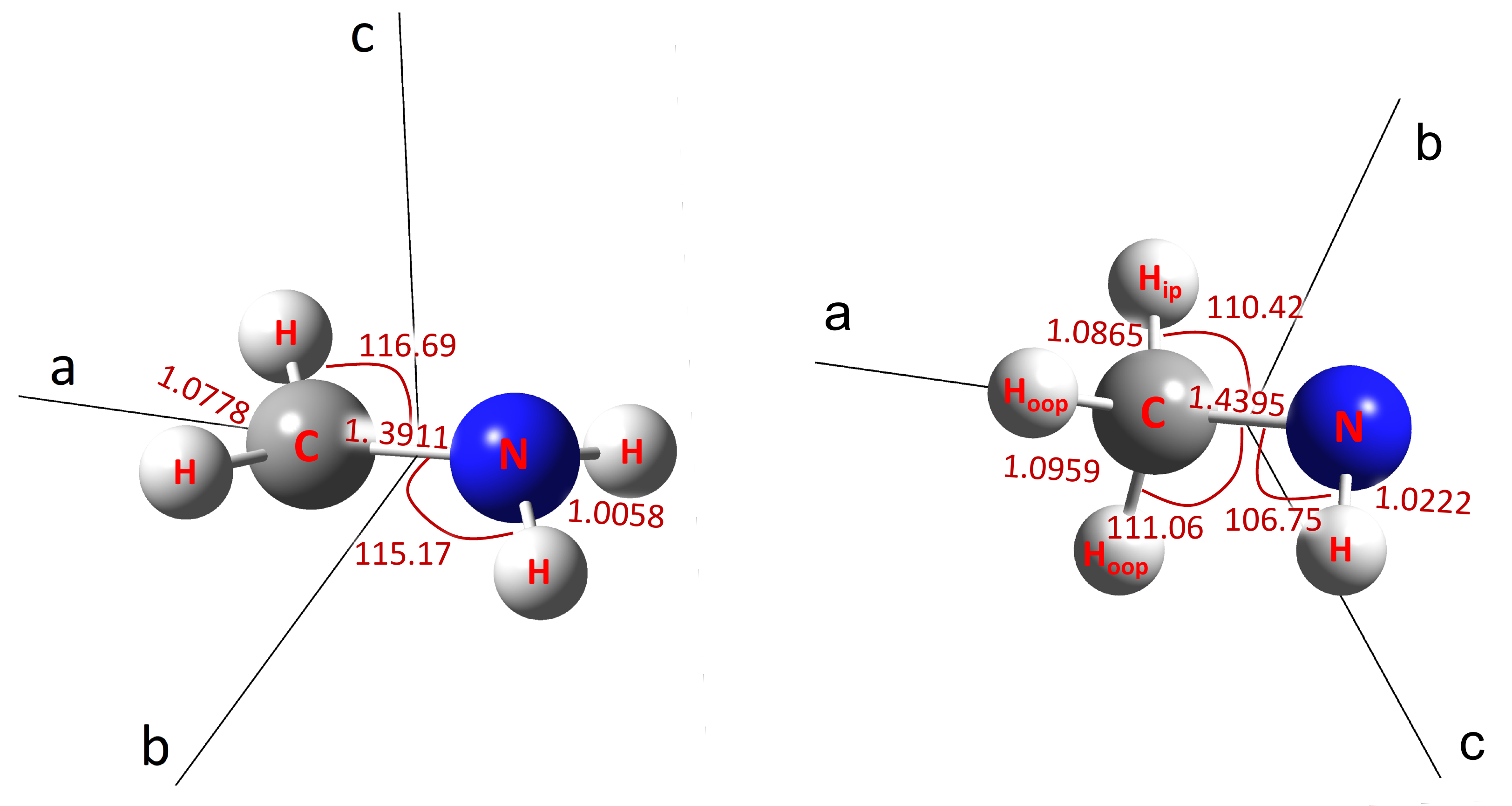}
\caption{Molecular structures of the CH$_2$NH$_2$ (left) and CH$_3$NH (right) radicals. ``Best-geo'' geometrical parameters (distances in \AA, angles in degrees) and inertial axes are also displayed. Dihedral angles: for CH$_2$NH$_2$; $\angle$HCNH = 39.47 deg., 171.16 deg; for CH$_3$NH, $\angle$H$_{oop}$CNH = $\pm$59.02 deg (oop stands for out-of-plane).
\label{fig-stru}}
\end{figure}

The molecular structures, together with some selected geometrical parameters, of the CH$_2$NH$_2$ and CH$_3$NH radicals are shown in Figure~\ref{fig-stru}. As mentioned in the computational details section, the composite approach denoted as ``best-geo'' scheme (see Appendix \ref{appendix:spectro}) has been employed in order to obtain very accurate equilibrium structures, and thus accurate equilibrium rotational constants. Interestingly, the geometrical parameters at this level of theory deviate by less than 0.001 \AA\ for bond distances and less than 0.1 deg. for angles from the ``cheap'' counterparts (see Appendix \ref{appendix:cheap}).

\begin{figure}
\centering
\includegraphics[width=1.\columnwidth]{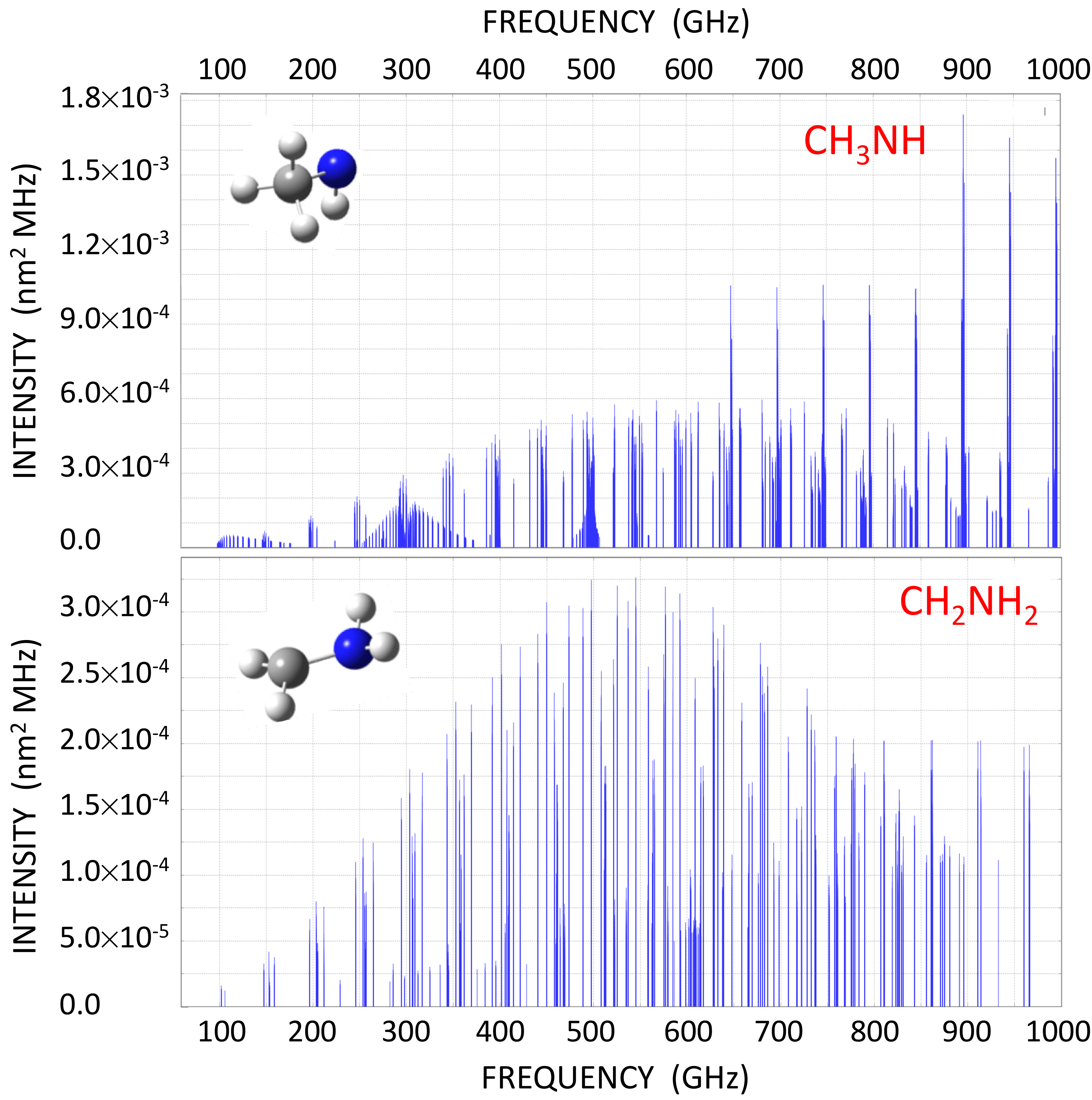}
\caption{Simulation of the rotational spectra of CH$_3$NH (top panel) and CH$_2$NH$_2$ (bottom panel) at $T$ = 100 K based on the spectroscopic parameters of Table~\ref{tab3}. 
\label{fig-spectra}}
\end{figure}

The list of spectroscopic parameters, computed as explained above and --in more details-- in the Appendix, is reported in Table~\ref{tab3}, with the principal inertia axes being displayed in Figure~\ref{fig-stru}. The spectroscopic properties of Table~\ref{tab3} have been employed to simulate the rotational spectra at $T$ = 100 K using the VMS-ROT software \citep{vms-rot}: the predicted rotational spectra in the 0-1000 GHz frequency range are depicted in Figure~\ref{fig-spectra}. 
According to the literature on this topic (see, e.g., \cite{rotstat,Puzzarini-IRPC2010_Review_rotational_spc,1.3503763,hso,1.4979573,benchS}), the rotational constants are expected to have an accuracy, in relative terms, of about 0.1\%, while the uncertainties affecting centrifugal-distortion constants and hyperfine parameters should not exceed 1-2\%. While these computational results do not have the required accuracy to directly guide astronomical searches, they can surely support laboratory experiments and their analysis (see, e.g., \cite{Puzzarini-IRPC2010_Review_rotational_spc,h2s36,proparg}). CH$_2$NH$_2$ and CH$_3$NH being radical species, the first challenge for a laboratory investigation is their {\it in situ} production. For this purpose, for example, electric discharge techniques \citep{hso,MELOSSO2019186} can be employed starting from methylamine as a precursor. 

\begin{table}
\caption{Computed spectroscopic parameters (in MHz) of CH$_2$NH$_2$ and CH$_3$NH.$^{a,b}$} \label{tab3}
\begin{tabular}{lc|lc} \hline  \hline      
\multicolumn{2}{c}{CH$_2$NH$_2$}   & \multicolumn{2}{c}{CH$_3$NH} \\  \hline 
$A_0$                      &   146501.69             & $A_0$                      &   124436.20 \\                   
$B_0$                      &   27393.55              & $B_0$                      &   25260.79  \\                   
$C_0$                      &   23642.74              & $C_0$                      &   24218.82  \\                
$\Delta_{J }$              &   4.85$\times$10$^{-2}$ & $\Delta_{J }$              &   5.58$\times$10$^{-2}$ \\       
$\Delta_{JK}$              &   2.88$\times$10$^{-1}$ & $\Delta_{JK}$              &   3.87$\times$10$^{-1}$ \\       
$\Delta_{K }$              &   2.73                  & $\Delta_{K }$              &   9.98$\times$10$^{-1}$ \\       
$\delta_{J }$              &  -6.82$\times$10$^{-3}$ & $\delta_{J }$              &   2.63$\times$10$^{-3}$ \\       
$\delta_{K }$              &  -3.06$\times$10$^{-1}$ & $\delta_{K }$              &  -1.76 \\                        
$\Phi_{J }$                &   2.57$\times$10$^{-8}$ & $\Phi_{J }$                &  -2.07$\times$10$^{-8}$ \\       
$\Phi_{JK}$                &   3.68$\times$10$^{-6}$ & $\Phi_{JK}$                &   1.09$\times$10$^{-4}$ \\       
$\Phi_{KJ}$                &   1.56$\times$10$^{-6}$ & $\Phi_{KJ}$                &  -3.79$\times$10$^{-4}$ \\       
$\Phi_{K }$                &   2.23$\times$10$^{-4}$ & $\Phi_{K }$                &   3.23$\times$10$^{-4}$ \\       
$\phi_{J }$                &  -8.53$\times$10$^{-9}$ & $\phi_{J }$                &   7.84$\times$10$^{-9}$ \\       
$\phi_{JK}$                &  -2.19$\times$10$^{-6}$ & $\phi_{JK}$                &   3.93$\times$10$^{-6}$ \\       
$\phi_{K }$                &  -1.19$\times$10$^{-4}$ & $\phi_{K }$                &   8.76$\times$10$^{-3}$ \\       
$\epsilon_{aa}$            &  -199.82               & $\epsilon_{aa}$            &  -1206.08               \\       
$\epsilon_{bb}$            &  -58.87                 & $\epsilon_{bb}$            &  -172.63                \\       
$\epsilon_{cc}$            &   6.409                 & $\epsilon_{cc}$            &   2.623                 \\       
$\widetilde{\epsilon}_{ab}$&   18.73                 & $\widetilde{\epsilon}_{ab}$&   338.61                \\       
$a_F$(N)                   &   12.95                 & $a_F$(N)                   &   33.62                 \\       
$T_{aa}$(N)                &  -8.41                  & $T_{aa}$(N)                &  -4.23                  \\       
$T_{bb}$(N)                &  -10.76                 & $T_{bb}$(N)                &  -4.23                  \\       
$T_{ac}$(N)                &  -7.96                  & $T_{ac}$(N)                &  -0.53                  \\       
$\chi_{aa}$(N)             &   1.84                  & $\chi_{aa}$(N)             &  -0.214                 \\       
$\chi_{bb}$(N)             &   0.180                 & $\chi_{bb}$(N)             &  -0.273                 \\       
$\chi_{ac}$(N)             &   1.05                  & $\chi_{ab}$(N)             &  -2.22                  \\       
$a_F$[H(N)]                &   7.19                  & $a_F$[H(N)]                &  -64.77                 \\       
$T_{aa}$[H(N)]             &  -1.28                  & $T_{aa}$[H(N)]             &  -39.78                 \\       
$T_{bb}$[H(N)]             &   6.43                  & $T_{bb}$[H(N)]             &   45.12                 \\       
$T_{ab}$[H(N)]             &   15.13                 & $T_{ab}$[H(N)]             &  -38.82                 \\       
$T_{ac}$[H(N)]             &   1.18                  &                            &                         \\       
$T_{bc}$[H(N)]             &   3.46                  &                            &                         \\       
$a_F$[H(C)]                &  -43.69                 & $a_F$[H(C-oop)]            &   127.50                \\       
$T_{aa}$[H(C)]             &  -20.85                 & $T_{aa}$[H(C-oop)]         &   6.48                  \\       
$T_{bb}$[H(C)]             &   17.13                 & $T_{bb}$[H(C-oop)]         &  -3.23                  \\       
$T_{ab}$[H(C)]             &  -27.10                 & $T_{ab}$[H(C-oop)]         &   4.45                  \\       
$T_{ac}$[H(C)]             &   0.78                  & $T_{ac}$[H(C-oop)]         &   5.88                  \\       
$T_{bc}$[H(C)]             &  -5.90                  & $T_{bc}$[H(C-oop)]         &   3.73                  \\       
                           &                         & $a_F$[H(C-ip)]             &  -2.00                  \\       
                           &                         & $T_{aa}$[H(C-ip)]          &   7.30                  \\       
                           &                         & $T_{bb}$[H(C-ip)]          &  -2.54                  \\       
                           &                         & $T_{ab}$[H(C-ip)]          &  -6.72                  \\ \hline
$\mu_a$ / D                &   0.931                 &$\mu_a$ / D                 &   1.246                 \\       
$\mu_c$ / D                &   0.504                 &$\mu_b$ / D                 &   1.472                 \\ \hline
 \end{tabular}
 
 \scriptsize{                                     $^a$ Watson A-reduction \citep{Wat77}.  ``oop'' stands for out-of-plane, ``ip'' for in-plane. See, Figure~\ref{fig-stru}. \\
 $^b$ Equilibrium ``best'' (CCSD(T)/CBS+CV+fT+fQ) rotational constants augmented by vibrational corrections at the B2PLYP-D3BJ/may$^{\prime}$-cc-pVTZ level. Quartic and sextic centrifugal-distortion constants at the B2PLYP-D3BJ/may$^{\prime}$-cc-pVTZ level. Equilibrium electron spin-rotation constants at the all-CCSD(T)/cc-pCVQZ level augmented by vibrational corrections at the B3LYP-D3(BJ)/6-31+G(d) level. Equilibrium values of Fermi-contact, anisotropic hyperfine coupling, and nuclear quadrupole coupling constants as well as dipole moment components at the all-CCSD(T)/aug-cc-pCVQZ(\_et5) level augmented by vibrational corrections at the B2PLYP-D3BJ/may$^{\prime}$-cc-pVTZ level. Anisotropic hyperfine and nuclear quadrupole tensors are traceless.
 } 
\end{table}

The rotational spectra displayed in Figure~\ref{fig-spectra} have been obtained considering all possible transitions with the rotational quantum number $J$ of the lower level ranging between 0 and 40. From the inspection of this figure, it is evident that both radicals show intense spectra, with their maxima shifting toward lower frequencies by decreasing the temperature and toward higher frequencies when increasing the temperature. As expected, the rotational spectra of the CH$_2$NH$_2$ and CH$_3$NH radicals are very different, but intense in both cases. According to Figure~\ref{fig-spectra}, in addition to possible difficulties in producing these radicals inside the spectrometer cell, the assignment of their spectra can be complicated by the fact that the most intense transitions lie well in the submillimeter-wave region. In fact, due to the propagation of the errors associated to the computed parameters when increasing the value of $J$, the uncertainties affecting the predicted transition values can be as large as 300-500 MHz (see, e.g., \cite{benchS}). However, one can rely on characteristic hyperfine pattern for helping the assignment procedure.

\section{Conclusions}

As mentioned in the Introduction, quantum-chemical calculations play a key role in the investigation of formation mechanisms in space because in many cases experimental studies are missing or even not feasible. Furthermore, the interpretation of the latter requires guidance of theory. In this respect, the CH$_3$NH$_2$ + CN reaction can be considered a paradigmatic example. Indeed, in \cite{C7CP05746F} the experimental work performed using the CRESU technique was supported by quantum-chemical calculations of limited accuracy combined with questionable interpretations of the latter, thus leading to the wrong conclusion that the product observed in the experiment was cyanamide. In a subsequent work \citep{acsearthspacechem.8b00098}, the quantum-chemical investigation was revised, thus pointing out the barrierless formation of CH$_2$NH$_2$ + HCN. However, in \citep{acsearthspacechem.8b00098}, the authors did not take the attack of CN to the NH$_2$ side into consideration. In the present study, we have taken a step further by investigating all possible reaction channels, thus demonstrating that two other pathways are feasible. A particularly important conclusion is that the reaction kinetics cannot be correctly described without the proper theoretical treatment of the barrierless entrance channels. Indeed, according to the results summarized in figure \ref{fig:rates}, P3 (CH$_3$NH + HCN) is the most favourable reaction product in the conditions considered. 

At very low temperatures, rates are exquisitely sensitive to energetics and kinetic barrier heights; therefore, high accuracy in quantum-chemical calculations can be a mandatory requirement in order to derive a correct picture. Indeed, even seemingly qualitative factors, whether reaction barriers following the formation of a pre-reactive complex lie above or below the initial reactants can fall within the uncertainty of the calculations, as demonstrated --for example-- in \cite{Vazart_JCTC2016_Formamide}. In this work, we have shown that two levels of theory commonly used in this field, namely the CBS-QB3 approach and the CCSD(T) method in conjunction with a triple-zeta quality basis set, are not suitable for quantitative results, especially when challenging open-shell species are involved. While the relative energies of stationary points are not strongly sensitive to the quality of the reference geometry, the situation is different for regions dominated by non-covalent interactions. For example, the presence of a barrier in the entrance channel for the formation of HCN + CH$_2$NH$_2$ claimed in \cite{C7CP05746F} is a computational artifact related to the well-known limits of the largely employed B3LYP functional. In this respect, the comparison with geometries issuing from accurate composite methods (here the ``cheap'' approach) and their impact on energy evaluations confirmed the effectiveness and reliability of the double-hybrid B2PLYP functional augmented by D3 dispersion corrections. 
   
To the best of our knowledge, the present investigation is the first one that has disclosed a feasible gas-phase pathway for AAN. On the other hand, this is hampered by the presence of competitive, more favorable, reaction channels, which make the formation of AAN unlikely to occur at extremely low temperature (e.g. 10-30 K). Nevertheless, in different environments, where there is an excess of energy, its feasibility cannot be excluded {\it a priori}. Concerning competitive reaction channels, the CH$_2$NH$_2$ and CH$_3$NH radicals, which are the most probable products, deserve to be spectroscopically characterized and might represent interesting intermediates toward further reactions. For these reasons, the rotational spectroscopic properties of these two radicals have also been computed with state-of-the-art methodologies. 

Finally, the accurate characterization of different gas-phase paths, of the corresponding stationary points by state-of-the-art quantum-chemical computations, and of the corresponding rate constants might provide useful pieces of information for building reliable chemical models for more complex networks.

\section*{Acknowledgements}
	
This work has been supported by MIUR ``PRIN 2015'' funds (Grant Number 2015F59J3R), by the University of Bologna (RFO funds) and by Scuola Normale Superiore (grant number SNS18\_B\_TASINATO). The SMART@SNS Laboratory (http://smart.sns.it) is acknowledged for providing high-performance computer facilities.

\bibliography{mnras_template}{}
\bibliographystyle{mnras}

\appendix

\section{Computational details}

In the following, the ``cheap'' geometry scheme as well as the CCSD(T)/CBS+CV and ``HEAT-like'' approaches are described in some details. Subsequently, the computational methodology for evaluating spectroscopic parameters is addressed.

\subsection{The ``cheap'' geometry scheme}
\label{appendix:cheap}

This composite scheme relies on the additivity approximation directly applied to the structural parameters (for more details, see \cite{ura,QUA25202}). Starting from the fc-CCSD(T)/cc-pVTZ optimized geometry, corrections to account for the basis set incompleteness as well as for core-valence correlation effect are introduced according to the following equation:

\begin{equation}
\begin{split}
r_{cheap}&=r \mathrm{(CCSD(T)/VTZ)} +\Delta r^{CBS} \mathrm{(MP2)} \\
&+ \Delta r_\mathrm{CV} \mathrm{(MP2)}  \; ,			
\end{split}
\end{equation}
where $r$ denotes a generic structural parameter. $\Delta r^{CBS} \mathrm{(MP2)}$ is the contribution stemming from the extrapolation to the CBS limit:

\begin{equation}
\begin{split}
\Delta r^{CBS} \mathrm{(MP2)} & = \frac{4^3 r\mathrm{(MP2/VQZ)} - 3^3 r\mathrm{(MP2/VTZ)}}{4^3 - 3^3} \\
&- r\mathrm{(MP2/VTZ)} \; ,
\end{split}
\end{equation}
which is obtained by extrapolating fc-MP2/VTZ ($n$=3) and fc-MP2/VQZ ($n$=4) calculations with the $n^{-3}$ extrapolation formula \citep{HKKN97}. 

The last term, $\Delta r_\mathrm{CV} \mathrm{(MP2)}$, is the core-valence (CV) correlation contribution, which is obtained as the difference between all electrons and fc MP2/cc-pCVTZ \citep{WD95} structural parameters. 

\subsection{The CCSD(T)/CBS+CV approach}
\label{appendix:cbscv}

CCSD(T)/CBS+CV denotes a composite scheme entirely based on CC theory to accurately evaluate the electronic energy of all the stationary points. CBS stands for complete basis set, thus meaning that CCSD(T) energies --obtained within the frozen-core approximation-- are extrapolated to the CBS limit. This extrapolation is performed in two steps. The CCSD(T) correlation contribution, extrapolated to the CBS limit by means of the $n^{-3}$ formula mentioned above \citep{HKKN97}:

\begin{equation}
\Delta E_\mathrm{corr}(n) = \Delta E_\mathrm{corr}^{CBS} +  A \, n^{-3} \; ,
\label{cbscc}
\end{equation}
is added to the HF-SCF CBS limit, evaluated by an exponential expression \citep{F93}:

\begin{equation}
E_\mathrm{SCF}(n) = E_\mathrm{SCF}^{CBS} + B \, exp \, (-C \, n) \; .
\label{cbshf}
\end{equation}
The cc-pVTZ and cc-pVQZ basis sets \citep{Dunning-JCP1989_cc-pVxZ} have been employed in the former equation, whereas the cc-pV$n$Z sets, with $n$=T,Q,5, have been used in the latter. 

By making use of the additivity approximation, the CV effects are taken into account by means of the corresponding correction:

\begin{equation}
\Delta E_\mathrm{CV} = E_\mathrm{core+val} - E_\mathrm{val} \; ,
\label{core}
\end{equation}
where $E_\mathrm{core+val}$ is the CCSD(T) total energy obtained by correlating all electrons and $E_\mathrm{val}$ is the CCSD(T) total energy computed within the fc approximation, both in the cc-pCVTZ basis set \citep{WD95}.

By putting together all these terms, the CCSD(T)/CBS+CV energy is obtained: 

\begin{equation}
E_{\mathrm{CBS+CV}}  = E_\mathrm{SCF}^{CBS} + \Delta E_\mathrm{corr}^{CBS}\mathrm{(CCSD(T))} + \Delta E_\mathrm{CV} \; .
\label{cbscv}
\end{equation}

\subsection{The ``HEAT-like'' approach}
\label{appendix:heat}

The reference for this approach is the HEAT protocol \citep{heat,heat2,heat3}, which has been reformulated as follows to provide the scheme denoted as ``HEAT-like'':

\begin{equation}
\begin{split}
E_{tot}  &=  E^{CBS}_{\mathrm{HF-SCF}} + \Delta E^{CBS}_{\mathrm{CCSD(T)}} + \Delta E_{\mathrm{CV}} + \Delta E_{\mathrm{fT}} \nonumber \\
	    &+  \Delta E_{\mathrm{pQ}} + \Delta E_{\mathrm{rel}} + \Delta E_{\mathrm{DBOC}}  \; ,
\end{split}
\label{heat}	   
\end{equation}
where the first three terms have been obtained as in the CCSD(T)/CBS+CV approach defined above. Corrections due to a full treatment of triples, $\Delta E_{\mathrm{fT}}$, and to a perturbative treatment of quadruples, $\Delta E_{\mathrm{pQ}}$, have computed --within the fc approximation-- as energy differences between CCSDT \citep{ccsdt1,ccsdt2,ccsdt3} and CCSD(T) and between CCSDT(Q) \citep{1.1950567,1.2121589,1.2988052} and CCSDT calculations employing the cc-pVTZ and cc-pVDZ basis sets, respectively. The diagonal Born-Oppenheimer correction, $\Delta E_{\mathrm{DBOC}}$  \citep{dboc1,dboc2,dboc3,dboc4}, and the scalar relativistic contribution to the energy, $\Delta E_{\mathrm{rel}}$ \citep{rel1,rel2}, have been computed at the HF-SCF/aug-cc-pVDZ \citep{KDH92} and CCSD(T)/aug-cc-pCVDZ (correlating all electrons) levels, respectively. The relativistic correction includes the (one-electron) Darwin and mass-velocity terms. 

\subsection{Spectroscopic characterization}
\label{appendix:spectro}

\subsubsection{The ``best-geo'' scheme}

The composite scheme employed for the determination of the equilibrium structure (and straightforwardly the equilibrium rotational constants) is a combination of gradient and geometry approaches. First of all, the CCSD(T)/CBS+CV equilibrium structure has been obtained by minimizing the following gradient:

\begin{equation}
\begin{split}
\label{eq1} 
\frac{dE_{\mathrm{CBS}}}{dx}& =
\frac{dE^{CBS}\mathrm{(HF-SCF)}}{dx} +\frac{d\Delta E^{CBS}\mathrm{(CCSD(T))}}{dx} \\
&+\frac{d\Delta E_{CV}}{dx} \; ,
\end{split}
\end{equation}
where the first two terms on the right-hand side are the energy gradients obtained using the extrapolation formula introduced in eqs. (\ref{cbshf}) and (\ref{cbscc}) for HF-SCF and the CCSD(T) correlation contribution, respectively. The aug-cc-pV$n$Z bases \citep{Dunning-JCP1989_cc-pVxZ,KDH92} have been employed, with $n$=T, Q and 5 being chosen for the HF-SCF extrapolation and $n$=T and Q being used for CCSD(T). Core-valence correlation effects have been considered in the gradient by
adding the corresponding correction, $d\Delta E_{CV}/dx$, where the energy difference is evaluated as in eq. (\ref{core}) and using the cc-pCVTZ basis set. 

Full triples and quadruples corrections have been obtained at the ``geometry'' level, by adding the following differences to the CCSD(T)/CBS+CV geometrical parameters:

\begin{equation}
\label{eq3} 
\Delta r(\mathrm{fT}) = r (\mathrm{CCSDT}) - r (\mathrm{CCSD(T)}) \; ,
\end{equation}
and
\begin{equation}
\label{eq4}
\Delta r(\mathrm{fQ}) = r (\mathrm{CCSDT(Q)}) - r (\mathrm{CCSDT}) \; ,
\end{equation}
where the cc-pVTZ basis set has been used for the fT correction and the cc-pVDZ set for the fQ contribution. This implies that geometry optimizations at the fc-CCSDT/cc-pVTZ, fc-CCSD(T)/cc-pVTZ, fc-CCSDTQ/cc-pVDZ, and fc-CCSDT/cc-pVDZ levels have been performed.

\subsubsection{Calculation of hyperfine parameters}

The electron spin-rotation tensor was calculated in a perturbative manner as second derivative of the energy with respect to the electron spin and rotational angular momentum as perturbations, as implemented in CFOUR and as described in \cite{jp103789x}. Since reduced Hamiltonians are actually used in the prediction or analysis of rotational spectra, for the off-diagonal term, the reduced value is provided, which has been determined as explained in \cite{BROWN1979111} and employing the computed vibrational ground-state rotational constants. Based on our previous experience \citep{hso}, the CCSD(T)/cc-pCVQZ level of theory should be able to provide accurate results that can quantitatively predict experiment.

The evaluation of the isotropic and anisotropic hyperfine coupling constants (hfcc) require the calculation of the spin density at the nucleus for the former and the corresponding dipole-dipole contributions for the latter (see, for example, \cite{1.466620}). Due to the importance of the effect of both very tight functions for
one-center terms and diffuse functions on the neighboring atoms (see, for example, \cite{1.466620,1.3503763,1.5128286}), CCSD(T) computations (with all electrons correlated) have been performed using the aug-cc-pCVQZ basis set for the C and N atoms, while a modified version (aug-cc-pCVQZ\_et5; \cite{1.3503763}) has been employed for hydrogens, which was obtained by adding five even-tempered uncontracted functions (for details, see \cite{1.3503763}).

Finally, as a byproduct of the hfcc calculations, the nitrogen quadrupole coupling constants have been obtained at the CCSD(T)/cc-pCVQZ level.

\subsection{CASPT2 potential for VRC-TST calculations}
\label{appendix:vrc}
VRC-TST calculations were performed computing energies on the dividing surfaces at the CASPT2 level. Since the number of sampling points necessary to reach the convergence threshold (5\%) in the Monte Carlo stochastic estimation of the reactive flux is rather large (tens of thousands of single point energy (SPE) evaluations are necessary for the investigated systems), it was decided, as it is customary for VRC-TST calculations, to determine SPEs using a relatively small active space and basis set, and then correct for the basis-set size, active space dimension, and geometry relaxation using a high level potential, which was parameterized as a function of the distance between the bond forming atoms. VRC-TST calculations for H abstraction from the methyl group were thus performed sampling the PES on the dividing surface using the cc-pVDZ basis set and a (5e,5o) active space that includes the unpaired electron orbital and the four $\pi$ and $\pi^*$ bonding and antibonding orbitals of the CN radical. The correction potential was determined performing relaxed geometry optimizations as a function of the distance between the abstracted H and the CN carbon atom in the 1.7-3.2~\AA~range using the cc-pVDZ basis set and a (7e,7o) active space equal to the (5e,5o) active space, with the addition of the $\sigma$ and $\sigma^*$ bonding and antibonding orbitals of the breaking \ce{C\bond{-}H} bond. Frequencies were also computed at the same level of theory. At the highest level of theory, the potential was computed on (7e,7o) relaxed geometries using the aug-cc-pVTZ basis set and a large (15e,13o) active space consisting of the (7e,7o) active space, with the addition of the $\sigma$ bonding and antibonding orbitals of CN (2e,2o), of the CN lone pair (2e,1o), of the C-N $\sigma$ and $\sigma^*$ bonding and antibonding orbitals of CH$_3$NH$_2$(2e,2o), and of the N lone pair of CH$_3$NH$_2$(2e,1o). In the case of CN addition to the CH$_3$NH$_2$ amino group to form the IC complex, VRC-TST calculations were performed using the cc-pVDZ basis set and a (7e,6o) active space consisting of the four $\pi$ and $\pi^*$ bonding and antibonding orbitals of the CN radical, of the unpaired electron orbital, and of the N lone pair of CH$_3$NH$_2$. As this active space includes all the orbitals expected to play a role in the formation of the IC complex, high level calculations were performed using the same active space and the aug-cc-pVTZ basis set. 

All CASPT2 calculations were performed using a 0.2 energy shift and the MOLPRO computational suite \citep{MOLPRO-WIREs,MOLPRO_brief}.

\bsp	
\label{lastpage}
\end{document}